\documentstyle [12pt]{article}

\begin{document}
\begin{titlepage}
\begin{flushright}
\noindent ITP 92--24\\
ZU-TH-38/92\\
\end{flushright}
\begin{center}
{\huge {\bf Charged Boson Stars }} \\
\vspace{0.2cm}
{\huge {\bf and }} \\
\vspace{0.2cm}
{\huge  {\bf Vacuum Instabilities}}\\
\vspace{0.2cm}
{\Large P. Jetzer}\footnote{Email address: K626230@CZHRZU1A.
Supported by the Swiss National Science Foundation.}\\
\begin{center}Institute of Theoretical Physics,
University of Z\"{u}rich, Sch\"onberggasse 9,
CH-8001 Z\"{u}rich, Switzerland\end{center}
{\Large P. Liljenberg}\footnote{Email address
address: tfepl@fy.chalmers.se.}
{\Large and B.-S. Skagerstam}\footnote{Email address
address: tfebss@secthf51.bitnet or tfebss@fy.chalmers.se. Research supported by
the Swedish National Research Council under contract no. 8244-103,
G\"{o}teborg.}\\
\begin{center}Institute of Theoretical Physics, Chalmers University
of Technology, S-412 96 G\"{o}teborg, Sweden\end{center}
\end{center}
\begin{abstract}
We consider charged boson stars and study their effect on the
structure of the vacuum. For very compact particle like
``stars", with constituent mass $m_{*}$ close to the Planck mass $m_{Pl}$,
i.e. $m_{*}^{2} = {\cal O}
(\alpha m_{Pl}^{2})$, we argue
that there is a limiting total electric charge $Z_c$, which,
primarily, is due to the formation of a pion condensate ($Z_{c} \simeq
0.5\alpha
^{-1}e$, where $\alpha$ is the
fine structure constant and $e$ is the electric charge of the positron). If the
 charge of the ``star" is larger than  $Z_c$ we find numerical evidence for a
 complete screening indicating a limiting charge for a very compact object.
 There
is also a less efficient competing charge screening mechanism
due to spontaneous
electron-positron pair creation in which case $Z_{c} \simeq \alpha
^{-1}e$. Astrophysical and
cosmological abundances of charged compact boson stars are
briefly discussed in terms of dark matter.
\end{abstract}
\end{titlepage}
\newpage
\setcounter{page}{1}
\begin{center} \section{\sc Introduction}\end{center}
Compact objects play an important role in current astrophysical research.
White dwarfs and neutron stars are examples of objects which
involve physics on scales down to the one of nuclei and
even of elementary particles. The recent developments in particle physics
and cosmology suggest that scalar fields may have played an important role in
the evolution of the early universe, for instance in primordial phase
transitions, and that they may make up part of the dark matter (for a recent
 account see e.g. Ref.\cite{graft}).
These facts motivated the study of gravitational equilibrium solutions
of scalar fields, in particular for massive complex fields, which form
so-called boson stars \cite{ph}.
Recently, compact objects made of charged bosons have been considered
and static spherically symmetric solutions were found \cite{phil1}.
For some charged boson star models their radius is  extremely small,
in which case a large electric charge can substantially modify the
structure of the vacuum.

In the presence of extended heavy nuclei the perturbative
vacuum of $QED$ is unstable if the number of charges $Z$ is larger than a
certain critical value $Z_{c}$. For $Z > Z_{c}$
there is spontaneous production of
electron-positron pairs \cite{zel} . For a positively (negatively)
supercritically charged and sufficiently compact object, as compared e.g. with
the Compton wavelength of the electron, pair-production is continued
until the created electrons (positrons)  shield the nucleus to an
effective charge $Z_{eff} \approx Z_{c}$ accompanied by the emission of
positrons (electrons) (for a review and detailed references on the subject
see e.g. \cite{raf}).
For a point-like charge it is, of course, well-known
that $Z\alpha > 1$ makes the Dirac Hamiltonian non-selfadjoint and the energy
eigenvalues become complex. For {\it extended} atomic nuclei the
limiting charge
has been estimated to be
close to $Z_{c} =173$. It has
been argued that if the size of the extended object tends to zero
$Z_{c}$ approaches $1/ \alpha$
\cite{gartner}.

For bosons and the Klein-Gordon equation
similar conclusions hold, however with a limiting value $Z\alpha \geq
0.5$, the equality being valid for the point-limit case (see
also in this context \cite{herb}).
As the charge of the source becomes larger than
the critical value $Z_c$ pairs of particles antiparticles (pions) will
be produced. The antiparticles (assumed to have the same sign for their charge
as the source) will be emitted at infinity, whereas the particles will
be tightly bound to the nucleus. Due to Bose statistics
a condensate with arbitrary many particles could be formed. However,
one also has to take into account the mutual repulsive Coulomb
interaction between the particles in the condensate. To add new particles
cost a certain amount of energy and therefore limits their number
in the condensate, which will screen the overcritical charge of the source
\cite{migdal}.

In the present paper we
confront these ideas about the existence of a limiting charge due
to the instability of the vacuum to the charged boson
stars. This might also be relevant for
the stability of the charged boson stars, which has
up to now been only investigated in terms of classical concepts \cite{phil2}.
A basic assumption  is the existence of stable and superheavy
charged scalar particles with mass $m^2_{*}={\cal O}(\alpha m^2_{Pl})$,
which could naturally appear owing to quantum fluctuations
in the very early phases of the universe, when its density was close
to $\rho \sim c^5/(G^2_{N} \hbar) = m_{Pl}^{4}$ \cite{markov}
(in natural units
$\hbar = c=1$ which we will use from now on).
It has been pointed out that the existence of very heavy
particles, called maximons in Ref. \cite{markov}, would unavoidably
lead to strong violation of thermodynamic equilibrium in the very
early universe \cite{sakharov}. A fact which is of importance for
the generation of a baryon asymmetry \cite{sakharov}.
If such fundamental particles exist, they
may form new compact structures, i.e. boson stars.

We will discuss two limiting cases of
the boson stars. In the case of a compact boson star, as compared to the
Compton wavelength of the electron, these ``stars" may have a typical size
comparable to the Planck length (!) and a mass of the order of Planck
mass or more. We will argue that these very compact objects
, which we shall still call boson ``stars", have a limiting total charge
close to $0.5/\alpha$. The time-scale $\tau$ of charge screening due to the
instability of the vacuum can be estimated to be of the same order as
for supercritical
atomic nuclei,
i.e. $\tau \leq {\cal O}(10^{-19})~s$ \cite{raf}. Expressed in terms
of the parameter
$m_{*}$ one may say $m_{*} >> (m_{*})_{cr}$, where $(m_{*})_{cr} =
m_{Pl}\sqrt{\alpha}$. In
the other limiting case $m_{*}$ will be
very
close to $(m_{*})_{cr}$. Under such a condition the star might even have a
macroscopic size
and the screening mechanism discussed above will be no longer efficient. These
considerations involve physics
both at the Planck scale as well as at the
scale of the electron or pion mass. The effective fine-structure constant will
then vary over this range of energy. This will, however, not change the
qualitative picture and thus we will neglect such quantum
effects.

In the
effective-potential approach to quantum resonances in stationary
geometries \cite{ruff1},
like the Kerr-Newman geometry,  the crossing of positive- and
negative- root classical solutions of the equations of motion for a test
particle with mass $m$ signals
particle-antiparticle production through the Klein process \cite{klein}.
The gap between particle and antiparticles
states, as described by the Dirac equation, is narrowed in the gravitational
field of a collapsed star, such as neutron stars. When the star collapses
to a black hole, the gap shrinks to zero at the Schwarzschild radius and the
vacuum becomes unstable
due to quantum tunneling \cite{ruff2}
in analogy with the particle production mechanism in strong electric fields
\cite{schwinger}.
Similarly one is led to a limiting charge-to-mass ratio for a
charged black-hole \cite{ruff2}. For the extended charged boson stars, which
we consider here, the
vacuum instability has a different character. In the effective-potential
language we will, e.g.,
see that classical positive- and negative-root solutions
will cross without a tunneling barrier unless the mass scale $m$ is much less
than $m_{*}$. Since the effective-potential language has been considered in
 quite a detail in this context, we will present some of these
 issues in the present paper.

It has been suggested that dark matter \cite{dark} may be composed of
charged massive particles, so called CHAMPs \cite{glas}. If the mass of
a CHAMP is of the order of Planck mass $m_{Pl}$, like {\it pyrgons}
\cite{kolb} or {\it maximons} \cite{markov},
one may wonder if gravitational binding effects may be important. As we will
argue, the compact boson "stars" can be considered to be a
self-consistent model for such superheavy  CHAMPs
gravitationally bound and classically stable particle-like objects.

The present paper should be regarded as preliminary in confronting
very compact charged boson stars with quantum mechanics.
Since these objects may
have a size comparable to the Planck length it is not clear what are the
effects of quantum gravity. We do not, however,
see a compelling reason why one should not pursue a study along the lines
indicated above for these objects,
as a first step towards a more detailed
understanding of the physics of these objects. In section 2 we describe the
basic features of charged boson stars. Various scenarios of vacuum
instabilities of very compact charged boson ``stars" are discussed in section
3. The effective potentials for fermions (bosons), i.e. $V_{eff}^{D\pm}$
($V_{eff}^{K\pm}$), are
derived in section 3.1 (section 3.2), where we also present a numerical
evaluation of them in the background fields of the compact ``star". We also
confront the physical picture as derived from these effective potentials with
the energy spectra directly derived from the Dirac or Klein-Gordon equation in
the external fields of the ``star". In section 3.2 we study in some
detail the
formation of a charged bose condensate close to the compact ``star" and
present the numerical evaluation of the coupled non-linear
equations of the bose condensate field and the electromagnetic field in the
background fields of the ``star". In the final section we discuss primordial
cosmological production of charged compact ``stars" by making use of
conventional kinetic equations. We argue that a background of such CHAMPs may
constitute at least a fraction of the dark matter of our universe.
\begin{center} \section{\sc Compact Charged Boson Stars}\end{center}
\setcounter{equation}{0}
The equations which describe the charged boson star are obtained from the
following action of a charged, massive boson field $\phi$ coupled to
gravity and a $U(1)$ gauge field, i.e.
\begin{equation}
S = \int d^{4}x\sqrt{-g} \left[-\frac{R}{16\pi G_{N}} +
g^{\mu \nu}(D_{\mu}\phi)^{*}(D_{\nu } \phi) - m^{2}_{*}|\phi |^{2} -
\frac{\lambda}{2}|\phi |^{4} - \frac{1}{4}F^{\mu \nu}F_{\mu \nu} \right]~~~,
\label{eq:1}
 \end{equation}
where
\begin{equation}
F_{\mu \nu} = \partial _{\mu} A_{\nu} - \partial _{\nu} A_{\mu}~~,~~
\end{equation}
and
\begin{equation}
D_{\mu} \phi = \partial _{\mu} \phi + ieA_{\mu}\phi ~~~,
\end{equation}
and $e>0$ is the electric charge of the positron.
 If the $U(1)$-gauge
symmetry is identified with the electromagnetic
field, $\alpha \equiv e^{2}/4\pi
  \approx 1/137$ is the fine
structure constant.
We will  consider the case $\lambda = 0$, but our results do not
crucially depend on the self-coupling term.
To the Lagrange density of Eq.~(\ref{eq:1}) one
may add a conformal coupling $R|\phi|^{2}/6$, where $R$ is
the scalar curvature, in
order to preserve conformal symmetry in the limit $m_{*} \rightarrow 0$
\cite{calcolja}. The inclusion of this conformal coupling has the effect of
``renormalizing" the  $|\phi |^{4}$-coupling but does not otherwise change the
equations of motion \cite{calcolja,krive}.
We therefore choose to disregard it.

A static spherically symmetric
and classically stable solution of the coupled non-linear differential
equations, as derived from the action Eq.~(\ref{eq:1}),
exists only if the gravitational attraction is larger than the Coulomb
repulsion, i.e. if $\alpha < e_{crit}^{2}$ , where $e_{crit}^{2}=
G_{N}m^{2}_{*}$. The space-time metric is of Schwarzschild form
 \begin{equation} ds^{} =
B(r)dt^{2} - A(r)dr^{2} - r^{2}(d\theta ^{2}
 + sin^{2}\theta d\varphi ^{2} )~~~,
\label{eq:2}
\end{equation}
and, furthermore,
\begin{equation}
A_{\mu } = (C_{}(r), 0,0,0)~~~,
{}~~~\phi (r,t) = \phi _{0} (r) e^{-i\omega t}~~~.
\label{eq:3}
\end{equation}
The eigenvalue $\omega$ corresponds physically
to the energy of the last charged constituent particle added
to the star.
The non-linear coupled differential equations for $ A(r), B(r),
{}~C(r)$ and $\phi_{0} (r)$, to be discussed in more detail in section 3.2,
 are solved numerically
with the boundary conditions
\begin{equation}
 A(0) = 1~~,~~ B(\infty) =0~~,~~dC(0)/dr =0~~,~~ C(\infty) = 0~~~,
\end{equation}
 and
\begin{equation}
 \phi_{0}(0) = \mbox{constant}~~,~~d\phi_{0}(0)/dr =0~~,~~\phi_0(\infty)=
d\phi_0(\infty)/dr=0~~~.
\end{equation}
These boundary conditions  describe a localized
object. As discussed in Ref.\cite{phil1}, the solution is most conveniently
described in terms of the rescaled
variables, i.e.
\begin{eqnarray}
r & \rightarrow & \tilde{r} = m_{*}r~~~,\nonumber \\
\phi _{0} & \rightarrow & \tilde{\phi} _{0} =
(8\pi G_{N})^{1/2}\phi _{0}~~~,\nonumber \\
C & \rightarrow & \tilde{C} = (\omega - eC)~~~,\nonumber \\
B & \rightarrow & \tilde{B} = m^{2}_{*}B~~~,\nonumber \\
e^{2} & \rightarrow & \tilde{e} ^{2}  =
\frac{e^{2}m_{Pl}^{2}}{8\pi m^{2}_{*}} =
 \frac{\alpha}{2}
\frac{m_{Pl}^{2}}{m^{2}_{*}}~~~,\nonumber \\
\lambda & \rightarrow & \tilde{\lambda} = \frac{\lambda m^{2}_{Pl}}{8\pi
m^{2}_{*}}~~~.
\label{eq:4}
\end{eqnarray}
In terms of the total mass $M$ and the particle number $N$ (or, equivalently,
the total charge $Q=eN$), the static solution has the asymptotic
Reissner-Nordstr\o m form, i.e. %
\begin{equation}
A(\tilde{r}) = (1 - \frac{2\tilde{M}}{\tilde{r}}
+\frac{2\tilde{N}^{2}\tilde{e}^{2}}{\tilde{r}^{2}})^{-1}~~~,
\label{eq:5}
\end{equation}
where $\tilde{M} = Mm_{*}/m^{2}_{Pl}$ and  $\tilde{N}=Nm^{2}_{*}/m^{2}_{Pl}$.
The radial component of the electric field has the asymptotic form
\begin{equation}
E(r) = -dC(r)/dr \approx Q/4\pi r^{2}~~~.
\end{equation}
The gravitational attraction overcomes the
Coulomb
repulsion if $\tilde{e}^{2} < 0.5$, i.e. $ m^{2}_{*} > \alpha m^{2}_{Pl} $. In
Fig.1 we exhibit a generic numerical solution for the coupled non-linear
equations of the metric functions $A, \tilde{B}$, the scalar field
$\tilde{\phi}_{0}$ and the
radial component of the electric field.
We notice that close
to the critical charge $\tilde{e}_{c}^{2} = 0.5$, or equivalently $m_{*} =
(m_{*})_{cr}$,
we obtain for a solution without nodes that \cite{phil1}
the following maximal values of $\tilde{N},\tilde{M}$ and $\tilde{R}$:
\begin{eqnarray}
 \tilde{N}_{max} & \approx  & \tilde{M}_{max} \approx 0.44 \cdot (\tilde{e}_{c}
-\tilde{e})^{-1/2}~~~, \nonumber \\
\tilde{R}_{max} & \approx &  1.5 \cdot (\tilde{e}_{c}
-\tilde{e})^{-1/2}~~~
\label{eq:6}
\end{eqnarray}
for $\tilde{\phi} _{0}(0)$ such that
\begin{equation}
\tilde{\phi} _{0}(0) =\tilde{\phi} _{c}(0)   \approx 0.0067 \cdot
(\tilde{e}_{c} -\tilde{e})^{1/2}~~~.
\end{equation}
Here $\tilde{R} = Rm_{*}$ is the rescaled mean radius
$R$ of the boson star, which is defined as follows
\begin{equation}
R = \frac{1}{eN} \int d^{3}xrJ^{0}~~~,
\label{eq:7}
\end{equation}
where
\begin{equation}
 J^{\mu} =  \sqrt{-g} g^{\mu \nu} \{ ie[\phi ^{*}\partial _{\nu} \phi -
\phi \partial _{\nu} \phi ^{*}] -2e^{2}A_{\nu}|\phi |^{2} \}
\end{equation}
is the conserved electromagnetic current.
If $\lambda \neq 0$ we obtain the following maximal values of
$\tilde{M}$ and $\tilde{R}$:
 \begin{eqnarray}
 \tilde{N}_{max} & \approx  & \tilde{M}_{max}
\approx  0.226 \cdot (\tilde{e}_{c}
-\tilde{e})^{-1/2} \frac{m_{Pl}}{m_{*}}\sqrt{\frac{\lambda}{8\pi}}~~~,
\nonumber
\\
\tilde{R}_{max} & \approx &  0.415 \cdot (\tilde{e}_{c}
-\tilde{e})^{-1/2}\frac{m_{Pl}}{m_{*}}\sqrt{\frac{\lambda}{8\pi}}~~~
\end{eqnarray}
for $\tilde{\phi} _{0}(0)$ such that
\begin{equation}
\tilde{\phi} _{0}(0) =\tilde{\phi} _{c}(0)   \approx 2.43 \cdot
(\tilde{e}_{c} -\tilde{e})^{1/2}~~~.
\end{equation}
We see that the scalar self-coupling does not change
the overall picture very much, as long as $m_{*} \sim {\cal O}(m_{Pl})$.
The mass-scale $m_{*}$ is therefore essentially our only
free parameter, if the $U(1)$-gauge
symmetry is identified with the electromagnetic field.

The dynamical stability of spherically symmetric charged boson stars has been
discussed in Ref.\cite{phil2}, where also the pulsation equation, which
determines the normal modes of the radial oscillations, has been derived.
The particle number $N$ and the mass $M$, as a
function e.g. of $\tilde{\phi}_{0}(0)$ or
equivalently the central density,
have their extrema, in particular their maximum, at the same value of
$\tilde{\phi}_{0}(0)$. From this fact
it follows that the pulsation equation has
 a zero
mode, where $N$ and $M$ have their extrema. It has been shown that for the
equilibrium solutions with a value of $\tilde{\phi}_{0}(0)$ bigger than a
certain critical value $\tilde{\phi}_{c}(0)$,
corresponding to the maximum mass, the pulsation equation has a negative
mode. Therefore, these configurations are dynamically unstable. On the other
hand for $\tilde{\phi}_{0}(0)$ less than $\tilde{\phi}_{c}(0)$ the equilibrium
solutions are stable.
Of course these results are based on a purely classical treatment of the
stability analysis. They do not take into account e.g. quantum effects, such
as
tunneling among different configurations with the emission of particles or
particle production in the strong electromagnetic and gravitational field
which are generated by the bose stars.

\begin{center}\section{\sc Decay
of the Vacuum}\end{center}
 \setcounter{equation}{0}
The properties of the vacuum in the presence of a charged
boson star are described by considering the spectrum of the
single-particle Dirac and Klein-Gordon equations in
the background of both gravitational and electromagnetic fields
of the star. We will restrict ourselves to the behaviour of the lowest
eigenvalue, corresponding to an $s-$wave.
When the bound state energy of the
electron dives into the negative-energy continuum the vacuum becomes
unstable and pair production occurs \cite{zel,raf}.
For charged bosons, as described by the
Klein-Gordon equation, the physics is quite different due to the formation of
a Bose condensate, see section 3.2 below.
\begin{center}
\subsection{\sc Vacuum Instabilities due to Fermions}
\end{center}
The Dirac equation describing a stationary state of an electron (with electric
 charge -$e$) with energy $E$ in the presence
of a static background gravitational field and a $U(1)$
gauge potential $A_{\mu }$, as given by
Eqs.(\ref{eq:2}) and (\ref{eq:3}), can, in a straightforward manner,
be reduced to
the following set of two coupled differential equations
\begin{eqnarray}
\frac{d}{dr}f(r) & = &\sqrt{A(r)}
\left( \frac{\kappa}{r} f(r) + m_{e}g(r)
-\frac{1}{\sqrt{B(r)}} \biggl( E+eC(r) \biggr) g(r)\right ) ~~~, \nonumber  \\
 \frac{d}{dr}g(r) & = &\sqrt{A(r)}
\left(m_{e}f(r) - \frac{\kappa}{r} g(r) +
\frac{1}{\sqrt{B(r)}}\biggl( E+eC(r) \biggr) f(r)\right) ~~~,
  \label{eqn:8}
\end{eqnarray}
where $\kappa = \pm n$  and $n$ is an integer.
We consider the ground state solution
for which $\kappa = -1$.
The functions $f$ and $g$ are the components of
the two-component spinor $\Psi$, which is defined as follows
\begin{equation}
\Psi(r,t) = \sqrt[4]{\frac{A(r)}{B(r)}} \left( \begin{array}{c}
f(r)\\
g(r)
\end{array}
\right)
e^{-iEt}~~~.
\label{eq:10}
\end{equation}
$\Psi$ is normalized in such a way that
\begin{equation}
\int_{0}^{\infty}dr \Psi ^{\dagger}(r,t)\Psi(r,t) = 1~~~.
\label{eq:9}
\end{equation}
The boundary conditions correspond to
normalizability and regularity at the origin, for which
$df(0)/dr=dg(0)/dr=0$.

It is clear from this equation that the natural length
scale for the wavefunctions
$f$ and $g$ is the Compton wavelength $m_{e}^{-1}$, where $m_{e}$ is the
electron mass.
As discussed in the previous section,
the mass of the particles forming the charged boson star is typically of
order $m_{Pl} \sqrt{\alpha}$ and their total number in the star is
${\cal O}(\alpha ^{-1})$.
For such stars the effective radius is roughly
${\cal O}(m_{Pl}^{-1})$,
which is much smaller than the Compton wavelength of the
electron. This situation is similar to the one we get for
pair production in the field of an extended nucleus, in which case  the
critical value $Z_{c}$ above which it occurs
is increased with respect to $1/\alpha$. $Z_{c}$ depends on the radius  of
the nucleus (see Ref.\cite{gartner}). For atomic nuclei $Z_{c} \simeq 173$.
In our case the radius of the boson star is several orders
of magnitude smaller than the
one of atomic nucleii and therefore we expect the value of $Z_{c}$ practically
to coincide with $1/\alpha$. In the pointlike limit, it turns out that
for $Z_c=1/\alpha$ the eigenvalues
of the Dirac equation become imaginary. A fact which
in our case will, however, not occur due to presence of a
finite radius, even if extremely tiny.
At the relevant scale, which is the
Compton wavelength of the electron, the metric
is practically flat and therefore we also do not expect any change
in the value of $Z_c$ induced
by the background gravitational field.

It has been argued \cite{ruff1,ruff2} that an effective potential can be used
to describe qualitatively the physics.  We can derive an effective potential by
 transforming  the Dirac  equation~(\ref{eqn:8}) into a
second order differential equation, i.e.
\begin{eqnarray}
&\mbox{}&\left( \sqrt{\frac{B(r)}{A(r)}}\frac{d}{dr}
+ \sqrt{B(r)}M(r) +  i\sigma _{2} (E-V(r))
\right)\times   \nonumber \\
&\mbox{}&\left( \sqrt{\frac{B(r)}{A(r)}}\frac{d}{dr}  -
\sqrt{B(r)}M(r)  - i\sigma _{2} (E-V(r)) \right)
\left(
 \begin{array}{c}
 f(r) \\
g(r)
\end{array}
 \right) = 0~~~,
\label{eq:square}
\end{eqnarray}
where the matrix $M(r)$ is given by
\begin{equation}
M(r) = \left(
\begin{array}{ll}
 \frac{\displaystyle \kappa }{ \displaystyle r} &
 m_{e} \\  m_{e} &
-\frac{\displaystyle \kappa }{\displaystyle r}  \end{array}
 \right) ~~~,
\end{equation}
and the potential $V(r)$ is
\begin{equation}
V(r) = -eC(r)~~~.
\end{equation}

If we neglect derivatives in $A$, $B$ and $V$, we can write
\begin{equation}
 - \frac{d^{2}}{d^{2}r_{*}} \left( \begin{array}{c}
f(r)\\
g(r)
\end{array}
\right)
 = (E-V^{D+}_{eff}(r))(E-V^{D-}_{eff}(r))
\left( \begin{array}{c}
f(r)\\
g(r)
\end{array}
\right)~~~,
\label{eq:deff}
\end{equation}
with
\begin{equation}
dr_{*}/dr=\sqrt{A(r)/B(r)}~~~.
\label{eq:38}
\end{equation}
Eq.~(\ref{eq:deff}) is in a suitable form for a WKB approximation.
Then the effective potential
$V^{D\pm}_{eff}$ for the Dirac equation is given by
\begin{equation}
\frac{V^{D\pm}_{eff}(r)}{m_{*}} = \frac{\tilde{C}(r) - \omega}{m_{*}}
 \pm \sqrt{B(r)} \left(
\frac{m^{2}_{e}}{m^{2}_{*}} +
 \frac{\kappa ^{2}}{\tilde{r}^{2}} \right)^{1/2}~~~.
\label{eqn:11}
\end{equation}
In Fig.2a
(2b) we exhibit $V^{D\pm}_{eff}$ as a function of $\tilde{r}$ for a positively
charged boson star with $m_{e}/m_{*} \simeq 0$ ($m_{e}/m_{*} = 1$). If the
 minimum of $V^{D+}_{eff}$ is less
than $-m_{e}$, pair production occurs for a positively charged boson star.
For dynamically stable charged boson equilibrium configurations
without nodes, which is the case for $\tilde{\phi}_{0}(0)$ less than
$\tilde{\phi}_c(0)$ corresponding to the maximal mass, we computed
$V^{D+}_{eff}$
and checked for which values of $\tilde e$ and $\tilde{\phi}_{0}(0)$ it
becomes less than $-m_e$ starting from some value on the $r$ axis.
(In practice since
$m_e^2/m_*^2 \ll 1$, it is accurate enough to see where $V^{D+}_{eff}$
becomes negative.) The negatively charged boson star can be studied in a
similar way.

It turns out that $V^{D+}_{eff}$ has negative values whenever the
value for the charge is nearby $Z=1/ \alpha$, and of course for
higher values of $Z$ (this can also bee seen by
making use of the asymptotic form
of Eq.~(\ref{eqn:11}) with $m_{e}/m_{*}\approx 0$, i.e. $V^{D+}_{eff}(r)
\approx
 (-\alpha N + |
\kappa |)/\tilde{r})$. Obviously since $V^{D+}_{eff}$ is an approximation the
so found value of $Z$ is nearby the expected exact one, which is just slightly
above $Z=1/\alpha$, to within an accuracy of about
10\%. In Fig.~3 we plotted together with the curves for the
charged boson star equilibrium
configuration also the borderline from which on
$V^{D+}_{eff}$ just starts having negative values.
We see that the dynamical stable equilibrium configurations, which lie
above this curve, are unstable against pair production. We expect
for them pair production to occur and the {\it additional}
charge above $Z_c=1/\alpha$ to be completely screened.

Since the size of the radius of the charged boson ``stars'' is extremely tiny
the electrostatic potential becomes very deep, such that
also heavier particles besides the electrons will be overcritical, leading
thus to their pair production. This effect, by considering
a nucleus with shrinking radius, has been studied in ref.
\cite{raf}, where the influence of the heavier leptons, like muons and
taus, has been taken into account.
It turns out that the overcritical nucleus,
with $Z > 1/\alpha$, gets sorrounded by shells of different leptons, but the
fact that the charge above $1/\alpha$ gets completely screened remains
unaffected.
Notice that to be completely selfconsistent
one should also take into account the fact that the heavier
leptons are not stable. Similar mechanism of pair production of
heavier particles will of course
also occur for charged overcritical boson ``stars''.
The net effect remains however that the charge above $Z_c=1/\alpha$ gets
completely screened, as mentioned above, and as long as such particles have
not masses of the order of the constituent particles itself we also do not
expect important back-reaction effects, which could alter the structure of the
star itself.

As an illustrative example we
have also computed the spectrum of the Dirac equation for
a fermionic particle of mass equal to $m_*$ rather
than the mass of the electron $m_e$.
In Fig.~4 we show $E/m_{*}$ as a
function of $\tilde{e}^{2}$, or equivalently $\alpha
m^{2}_{Pl}/2m^{2}_{*}$, for various values of $\tilde{\phi}_{0}(0)$ for the
$1s_{1/2}$ state ($\kappa = -1$). We see that in this case $E$ decreases by
increasing $\tilde e$ and can even become negative. However, not as much as
$E/m_*=-1$ and therefore pair production {\it does not} occur. This because
the mass $m_*$ is too heavy. For this case the presence of gravity plays an
important role, since now the typical length scale of the  wavefunctions
$f$ and $g$ is of order $1/m_*$, a distance at which the metric is still
far from being flat. The solution of the Dirac equation in the external fields
of the compact charged boson ``star" leads to a physical picture in agreement
with
the one obtained by making use of the effective potential $V^{D\pm}_{eff}$.

\begin{center}
 \subsection{\sc Vacuum Instabilities due to Bosons}
\end{center}
One may also consider pair production of
charged bose particles, like for instance
pions.
For atomic nuclei at normal nuclear density the critical value $Z_c$
above which pion pair production occurs is large: $Z_c \sim 3000$
\cite{raf}. As for fermions it depends on the radius of the nucleus.
It decreases by decreasing the nuclear size. Since the charged boson stars
are extremely small the corresponding value $Z_c$ is much smaller and is
actually very close to $0.5/\alpha$ \cite{bawin}. In Fig.5 we illustrate this
 approach to the critical charge $Z_{c}$ by considering the ground state
energy $E(Z,R)$ of the Klein-Gordon equation in the external field of an
extended uniformly charged sphere with radius $R$
 (in Ref.\cite{bawin} one considers for a radius smaller
 than $R$ a constant potential).
Analytic solution can be obtained for $r \le R$ and $r\geq R$.
 These solutions are then numerically linked together at $r=R$ in a standard
manner \cite{popov}. We notice that $\partial E(Z,R)/\partial Z$ tends to minus
infinity as $R$ tends to zero for $Z$ close to $Z_{c}$.

Once the lowest bound state of the Klein-Gordon equation dives into the
continuum pair production of pions occurs.
Due to Bose statistics a condensate of pions is then formed. The number of
pions in the condensate can be quite large. It is limited, however, due to
the Coulomb repulsion among the pions, such that above a certain critical
number it is energetically no more possible to add new pions to the
condensate \cite{raf}.

The presence of such a bose condensate, which  is described
by the negatively charged scalar field $\eta$ of the pion or any other
similar charged scalar field, can be conveniently
incorporated by including an additional
term $S_{con}$ to the
action Eq.~(\ref{eq:1}), i.e.
\begin{equation} S_{con} = \int d^{4}x\sqrt{ -g} \left[ g^{\mu \nu}
(D_{\mu}\eta)^{*}D_{\nu}\eta - m^{2}|\eta |^{2} \right]~~~,
\label{eq:p1}
\end{equation}
where now
\begin{equation}
D_{\mu} = \partial _{\mu} -ieA_{\mu}~~~.
\end{equation}
Higher
order terms in the field $\eta$ can be added to this action
if there are additional interactions among the $\eta$ fields as
for instance a $\lambda \eta^4$ term \cite{migdal}. Here, however, for clarity
we restrict ourselves to this form. The electromagnetic and
gravitational fields couple to both $\eta$ and $\phi$.
Variation of the various fields leads now to the following
equations of
motion, i.e. the two Einstein equations
\begin{eqnarray}
\frac{A'}{A^{2}r} +\frac{1}{r^{2}}(1-\frac{1}{A}) & = &
8\pi G_{N} \left[ (\frac{(\omega - eC)^{2}}{B}
 + m^{2}_{*})
 \phi_{0}^{2} \right. + (\frac{(E + eC)^{2}}{B} +  m^{2})\eta _{0}^{2}
\nonumber \\
& + & \left.
\frac{\phi_{0}^{'2}}{A} + \frac{\eta_{0}^{'2}}{A}
+ \frac{C^{'2}}{2AB} \right]~~~,
\label{eq:p2}
 \end{eqnarray}
and
\begin{eqnarray} \frac{B'}{ABr} -\frac{1}{r^{2}}(1-\frac{1}{A}) & = &
8\pi G_{N} \left[ ( \frac{(\omega - eC)^{2}}{B} - m^{2}_{*})
 \phi_{0}^{2} + (\frac{(E + eC)^{2}}{B} -  m^{2})\eta_{0}^{2} \right.
\nonumber \\
& + & \left.
\frac{\phi_{0}^{'2}}{A} + \frac{\eta_{0}^{'2}}{A}
- \frac{C^{'2}}{2AB} \right]~~~,
\label{eq:p3}
\end{eqnarray}
the Maxwell equation
\begin{equation}
C'' + (\frac{2}{r} -\frac{A^{'}}{2A} -\frac{B^{'}}{2B})C^{'} + 2e\phi
_{0}^{2}A(\omega - eC) - (E + eC)2eA\eta_{0}^{2} = 0~~~,
\label{eq:p4}
\end{equation}
and the scalar wave equations for $\phi_0$
\begin{equation}
\phi _{0}'' + (\frac{2}{r} -\frac{A^{'}}{2A}
+\frac{B^{'}}{2B})\phi^{'}_{0} +
A(\frac{(\omega - eC)^{2}}{B} - m^{2}_{*})
\phi _{0} = 0~~~,
\label{eq:p5}
\end{equation}
where a prime denotes differentiation with respect to $r$.
For the field $\eta$ we have assumed
\begin{equation}
\eta(r,t)=\eta_0(r) e^{-iEt}~~~,
\end{equation}
and hence we also obtain
\begin{equation}
\eta ^{''}_{0} + (\frac{2}{r} -\frac{A^{'}}{2A}
+\frac{B^{'}}{2B})\eta^{'}_{0} +
A(\frac{ (E + eC)^{2}}{B} - m^{2})
\eta _{0} = 0~~~.
 \label{eq:p6}
\end{equation}
In order to solve this Klein-Gordon equation we impose the boundary conditions
\begin{equation}
\eta_0(0)=const~~~,~~~d\eta_0(0)/dr=0~~~,~~~
\eta_0(\infty)=d\eta_0(\infty)/dr=0~~.
\label{bc}
\end{equation}

Similarly to the Dirac case one can also
find an effective potential $V^{K\pm}_{eff}$. If we neglect derivatives of
the metric functions $A$ and $B$, the Klein-Gordon equation (\ref{eq:p6})
leads to
\begin{equation}
-\frac{d^{2}u(r)}{d^{2}r_{*}} = (E-V^{K+}_{eff}(r))(E-V^{K-}_{eff}(r))
u(r)~~~,
\end{equation}
where
\begin{equation}
\eta_0(r)=u(r)/r~~~,
\end{equation}
and $dr_*$ is defined as in eq.(\ref{eq:38}).
 The effective potential for the Klein-Gordon
equation is then given by
 \begin{equation}
 \frac{V^{K\pm}_{eff}(r)}{m_*}=\frac{\tilde C (r) -\omega}{m_*}\pm
\sqrt{B(r)}\frac{m}{m_*}~~,
 \label{eq:13}
 \end{equation}
where $\tilde C (r)$ is defined in
Eq.(\ref{eq:4}).
We found, however, that it is difficult to numerically determine the
critical charge with hig accuracy by making use of this effective potential.
We notice that  $V^{K\pm}_{eff} = V^{D\pm}_{eff}|_{\kappa = 0}$
\footnote{The two effective potentials
are equal as far as non-relativistic
effects are concerned. This can be seen by making use of the
the so called Langer modification of the
effective potential \cite{langer}: we perform the
shift $m/m_{*} \rightarrow \sqrt{(m/m_{*})^{2} +(l+1/2)^{2}/r^2}$ in
$V^{K\pm}_{eff}$, where $l$ is the angular momentum quantum number, and in
$V^{D\pm}_{eff}$ the shift $\kappa \rightarrow \kappa + 1/2$.
$V^{K+}_{eff}$
and $V^{D+}_{eff}$ lead to the correct spectrum in the
non-relativistic limit by making use of the WKB-approximation.
With the Langer modification
$V^{K+}_{eff}$ will change sign for some value of $\tilde{r}$ at
$Z=Z_{c} = 0.5/\alpha$ within an accuracy of about $15\%$ if $m/m_{*} \ll 1$.}.
In Fig.6a (6b)
we exhibit  $V^{K\pm}_{eff}(\tilde r)$ for a
positively charged boson star with
$m/m_{*} = 0.05$ ($m/m_{*} = 1$). In Fig.6a we see that the bound state of the
 negatively charged particle dives into the negative continuum, corresponding
 to anti-particle states, without the presence of a tunneling barrier. If
$m/m_{*} = 1$ there is a finite energy barrier between the spectrum of particle
and anti-particle.

We investigate now the effect of a pion condensate on a charged boson
star in which case $ m^{2}/m^{2}_{*} = {\cal O} (m^{2}_{\pi}/\alpha
m^{2}_{Pl})$, which, of course, is a very small number.  In our case the
compact object can therefore be considered as a  charged point source. In fact
the typical length scale of the pion condensate is the Compton wavelength of a
pion, i.e.  the  compact object
has a size much smaller than
the bose condensate as follows from Eq.(\ref{eq:p6}). Therefore, the fields
$\phi _{0}(r)$, $A(r)$ and $B(r)$, as described by
the equations of motion
Eqs.(\ref{eq:p1})-(\ref{eq:p6}),
will be very localized as compared to
variations of the fields $\eta_0(r)$ and $C(r)$ (see Fig.1a and 1b). It is
 reasonable
to assume
that under such circumstances we can neglect the effect of the bose
condensate on the compact object itself.

We checked this by solving numerically the full set of
Eqs.(\ref{eq:p2})-(\ref{eq:p6}) for some values of the ratio $m/m_{*}$.
It turns out that the back-reaction of the condensate
on the boson star itself is small even for relatively
large mass ratios. In fact the total mass, particle number and size of
the star is very little changed by the presence of the condensate.
We illustrate this feature for a
dynamically stable charged
boson star with $\tilde{e}^{2} = 0.25$ and $\tilde{\phi} _{0}(0) = 0.2$, which
corresponds to about $N \approx 66$ particles for the compact object
itself when there is no condensate present. If we now include a condensate with
a mass
ratio $m^{2}/m^{2}_{*} = 1/1000$, we
find that the boson star remains
essentially unaffected (the particle number as well as its mass changes only
within less than one percent) apart from its electromagnetic
properties at large distances. As seen in Fig.7, the charge distribution of the
condensate is
localized mostly outside the compact object and contains $N_{c} \approx 10$
particles, i.e. at large distances the star, however, appears to have
a net
charge $Q \approx 56e$.
We have verified numerically that the back-reaction of
the condensate on the boson star
gets even less when $m^{2}/m^{2}_{*}$ decreases.

We now return to the issue concerning a pion condensate on a
charged boson star, in which case $m$ is the mass of the pion.
At distances much
larger as compared
to $m_{*}^{-1}$ we therefore regard the term in Maxwell's equation
\begin{equation}
\rho_{ext}(r) = 2e\phi _{0}(r)^{2}A(r)(\omega - eC(r))
\label{eq:p7}
\end{equation}
as a localized external charge distribution for which we can
put $A(r) \approx 1$.
Let $C_0$ be a solution to
\begin{equation} -\left(C_{0}'' + \frac{2}{r}C_{0}'\right)=
\rho_{ext}~~~,  \label{eq:p10}
\end{equation}
i.e. at large distances
\begin{equation}
C_{0} =  \frac{Q}{4\pi r}~~~.
\end{equation}
We now write $C = C_{0} + C_{b}$.
At  distances which are large compared to $m^{-1}_{*}$, Maxwell's
equation becomes
\begin{equation}
 \left(\tilde{C}_{b}'' + \frac{2}{r}\tilde{C}_{b}'\right)= \left( E +
 eC_{0 }+ eq\tilde{C}_{b}  \right)2e\tilde{\eta} ^{2}_{0}
\label{eq:p8}~~,
\end{equation}
and the Klein-Gordon equation
\begin{equation}
\tilde{\eta} _{0}^{''} + \frac{2}{r}\tilde{\eta} _{0}^{'} +
\left( (E+ eC_{0} + eq\tilde{C}_{b})^{2} - m^{2} \right) \tilde{\eta} _{0}  =
0~~~.
 \label{eq:p9}
\end{equation}
Here we rescaled the fields as follows:
\begin {equation}
 \tilde{C}_{b}=C_{b}/q~~~,~~~\tilde{\eta} _{0} = \eta _{0}/\sqrt{q}~~~,
\end{equation}
 where $q$ is choosen in such a way
that
\begin{equation}
2e^2\int d^{3}x~\tilde{\eta}^{2}_{0}\left( E + eC_{0} +eq\tilde{C}_{b}
\right)
=1~~~. \label{eq:pp10}
\end{equation}
The $\tilde{\eta}_{0}$-field is
normalized to a unit charge. The parameter $-q$
corresponds thus to the total electric charge of
the condensate.
The physical interpretation of these equations is now clear.
The set of eqs (\ref{eq:p9}) and (\ref{eq:pp10}) is analogous to the
Thomas-Fermi equation, which describes in a self-consistent way the
electron wave functions in the field of a nucleus. The repulsive interaction
between the pions is crucial to ensure the stability of the system.
Otherwise there would be production of an arbitrarily large number
of particle antiparticle pairs.

The eigenvalue $E$
corresponds to the energy of the last particle added to the condensate
\cite{raf}. A natural choice is $E= - m$, where $m$ is the rest mass of the
pion. With this choice $\tilde\eta_0$ does no longer decrease exponentially
for $r \to \infty$, as can be seen from eq. (\ref{eq:p9}). As a result
it turns out that the
normalization integral, eq. (\ref{eq:pp10}), is divergent.
This fact may indicate that the mean field approximation breaks down, or
that one has to take into account other interactions among the pions,
as for instance a $\lambda\eta^4$ term.

However, if weak decays of the pions are taken into account the
first source of instability is the ``inverse $\beta$ decay'' \cite{bawin}:
$N \to (N,\pi^-) + e^+ +\nu_e$. $N$ denotes the ``star'' (positively charged)
and $(N,\pi^-)$ a $\pi^-$ bound state in the field of the ``star''.
This process sets in for $Z \ge Z_c$, as soon as the ground state energy of
the Klein-Gordon equation is lower than $-m_e$
(the energy of $\nu_e$ can be arbitrarily small), rather than $-m$.
As we have seen above when the radius
of the ``star"
is sufficiently small compared to the Compton wavelength of an electron or a
pion and $Z \ge Z_c$, the ground state energy of the Klein-Gordon
equation becomes very rapidly negative, see Fig.5.
In this case one can set $E=-m_e$ in eq.(\ref{eq:p9}). Then $\tilde\eta_0$
falls off exponentially for $r \to \infty$ and thus
the integral (\ref{eq:pp10}) is finite.

We have numerically solved the above Eqs.(\ref{eq:p8})-(\ref{eq:pp10})
with $E=-m_e$ for the electromagnetic field $\tilde C_b$ and
the condensate $\tilde{\eta}_{0}$ using an iterative method as
discussed in Ref.\cite{raf}. For our purposes it is sufficient to consider a
uniformly positively charged solid sphere with radius $R$. The critical charge
$Z_{c}$ is first determined by solving the corresponding Klein-Gordon equation
with $q=0$ for a
negatively charged particle with binding energy $E$ close to zero. The
corresponding solution is then inserted into
Eq.(\ref{eq:p8}) as  initial data for
$\tilde{\eta}_{0}$ in the iterative method, where we now solve for
 $\tilde{C}_{b}$. With it
we solve again for $\tilde{\eta}_{0}$
in Eq.(\ref{eq:p9}) with $q$ as an eigenvalue and renormalize
 the solution according to
Eq.(\ref{eq:pp10}).
This iterative procedure
converges rapidly and it is interrupted
when the desired numerical accuracy in $q$ is
reached. In Fig.8 we present the result of such a numerical calculation for
 three
different values of $R$ in terms of an effective charge $Z_{eff} = Z-q$ of the
``star".
 It turns out that, as already discussed to some extent in
Ref.\cite{raf}, the number of particles $q/e$ in the charged condensate
depends on $\delta Z$ ($\delta Z=Z-Z_c$ is the amount of
charge above the critical value) and also on the radius $R$ of the ``star''.
For very compact objects, with a radius smaller than pion's Compton
wavelength, our numerical
calculations show that the absolute value of $q$ is bigger than $\delta Z$.
This tendency increases by shrinking the radius.
It may even turn out that in the point-like limit, for
$Z \geq 0.5/\alpha$, the pion condensate will completely
screen the object, i.e. it becomes neutral.
We would therefore obtain a limiting
charge $Z_{c} = 0.5/\alpha$ for a point-like charge defined as the limit of an
extended charge distribution. Recently Gribov and Nyiri \cite{gribov} have
reached a similar conclusion, however in the approximation
of a massless ``pion''. We
intend to return to this issue elsewhere.

As already mentioned in the previous section on fermion instabilities, due
to the extremely small radius of the ``star'' also particles heavier than
the pions could contribute to the condensate. Like in the
fermion case, we do not expect this fact to alter significantly the
result for the screening of the overcritical charge, since this
depend primarily from the electromagnetic charges involved and not from the
the particle mass. As mentioned above we solved numerically the full
set of eqs. (\ref{eq:p8}) - (\ref{eq:pp10}) for large values of the
the ratio $m/m_*$ (in particular for $\approx$ 1/32) in order to study possible
back-reaction effects on the star itself. We find that such effects are
negligible. For larger $m$, comparable to $m_*$, the vacuum is however
no longer overcritical (see Fig. 4) and thus such particle will not
be produced. We conclude thus that within our Thomas-Fermi like approach
back-reactions are  not important. Of course the actual composition of the
condensate may change from the simple one particle-type solution
presented here, but not the net screening effect. This last point is what is
most important for our following astrophysical applications.

In this section we have shown that the vacuum is unstable against pair
 production of fermions or bosons if the charged bose-``star" is
 overcritical, i.e. if $Z \ge Z_{c}$. The time-scale, $t_{c}$, of the
 destabilization of the vacuum can be obtained from the structure of the bound
state wave function of the Dirac equation or Klein-Gordon equation close to the
edge of the negative continuum, the negative continuum states themselves and
 the structure of the potential $C$ in terms of an overlap integral of these
 quantities, as far as one considers only one state diving into the negative
 continuum (c.f. chapter 6 in Ref.\cite{raf}). The natural
scale determining the relevant time-scale,
therefore, corresponds to the scale of the bound state wave functions, i.e. the
 electron mass or the pion mass. For our purposes $t_{c}$ can be estimated,
 within a few orders of magnitude,
by making use of the probability density per unit time, $\omega$, for pair
 production of particles with mass $m$ and spin $s$ in a strong
electric field $E$
 \cite{schwinger}, i.e. ($\hbar=c=1$)
\begin{equation}
\omega = (2s+1)\frac{\alpha E^{2}}{2\pi ^{2}} \sum _{n=1}^{\infty}
\frac{1}{n^{2}}(-1)^{(2s+1)(n+1)}\exp \left( -\frac{n\pi m^{2}}{eE} \right)~~~,
\end{equation}
where we use $eE \simeq N\alpha/r_{C}^{2}$ and $r_{C} = 1/m$. We can then
 estimate the change in the screening charge, $dQ/dt$, by writing
(c.f. chapter 21 in Ref.\cite{raf})
\begin{equation}
\frac{1}{e}\frac{dQ}{dt} \simeq r_{C}^{3}\omega~~~~.
\end{equation}
The time-scale $t_{c}$ for a process for which the change in particle number
is of order one, i.e.
$\Delta Q/e \simeq 1$, is
\begin{equation}
t_{c} \simeq  \frac{8\pi ^{3}r_{C}}{\alpha ^{2} N^{2}(2s+1)A}~~~,
\end{equation}
where $A \approx 0.022 (\approx 0.021)$ for fermions (bosons) and
we used $N \simeq 1/ \alpha $.
For the charged and very compact bose ``stars"
we thus obtain $t_c \simeq 10^{-17} s$ for the
pair production of electrons and positrons, whereas
for pions we get a time-scale which is
three orders of magnitude smaller, i.e. $t_c \simeq 10^{-20} s$. From
these values we conclude that the pion condensate
most likely will form first and is thus an
efficient screening mechanism once the radius of the bose  ``star'' is
less than $0.1 fermi$.
\begin{center} \section{\sc
Final Remarks}\end{center}
\setcounter{equation}{0}
The very compact charged boson
``stars" taken as a self-consistent model for CHAMPs
involve physics at Planck scale, a fact this which
makes it difficult to perform reliable
estimates of their possibile cosmological relic density.
Nevertheless, below we
 suggest some
plausible arguments in favour of such a relic abundance which may even
 be observable and make up a substantial fraction of the dark matter
present in the universe.

A bound on the mass $M$ of CHAMPs is found
by considering their cosmic relic
 number density,
$n _{M}$, along the lines discussed in Ref.\cite{glas}.
We assume that CHAMPs, in the form of compact boson ``stars"  with mass $M$,
were formed in the very early universe.
At a
temperature $T = T_{*}$, such that $M/T_{*} \simeq 40$ \cite{wolfram},
the superheavy CHAMPs (and anti-CHAMPs) will freeze
out due to the equality between the Hubble
expansion rate and the annihilation rate of particles and anti-particles.
Then, a bound on the mass $M$
emerges by equating the relic energy density to the critical density
$\rho _{c} = 3H^{2}_{0}/8\pi G_{N} \approx 10^{-29}gcm^{-3}$, where
$H_{0} \approx 100kms^{-1}Mpc^{-1}$ is the Hubble parameter, i.e.  we assume
 that the
superheavy
CHAMPs constitute the dark matter of the universe \cite{dark}.
If only
electromagnetic interactions are taken into account,
we would get $M \simeq {\cal
 O}(TeV)$.
For larger values
of $M$ the universe becomes matter dominated. Such an estimate is based on the
 assumption that the CHAMPs are in
{\it thermal equilibrium} at sufficiently high
 temperatures as compared to their rest mass. Since in our case $M \simeq
 {\cal O}(1/\sqrt{\alpha}) m_{Pl}$, this may be a doubtful assumption.

One may instead assume that CHAMPs are thermally produced in the early
 universe starting with
a very small or even a vanishing initial density. This way we
get a {\it lower} bound on their
 present relic abundance for a given ratio $M/T_i$. $T_i$ is the
temperature where the termal production is assumed to begin.
The thermal production (and annihilation) in an
expanding universe is assumed, in analogue with thermal production of grand
 unified magnetic monopoles \cite{turner}, to be described by a Boltzmann
 equation
\begin{equation}
\frac{df(x)}{dx} = Z\left( f^{2}(x) -g(x)f^{2}_{\gamma}(x)\right)~~~,
\label{eq:boltz}
\end{equation}
where
\begin{equation}
x= T/M~~,~~f(x) = n_{M}(T)/T^{3}~~,~~f_{\gamma}(x) = n_{\gamma}(T)/T^{3}~~~.
\end{equation}
 Here
$n_{\gamma}(T)$ is the number density of photons at temperature $T$. Since we
consider a radiation dominated phase for the
early universe, we have the following relation between the temperature $T(t)$
and the time $t$:
\begin{equation}
T(t)^{2} = \sqrt \frac{45}{16\pi ^{3}G_{N}N_{eff}}~ \frac{1}{t}~~~,
\end{equation}
where $N_{eff} \simeq 427/4$ is the effective number of degrees of
freedom for the standard model well above the electro-weak scale. We also have
that $R(t)T(t)$ remains constant in time,
where $R(t)$ is the cosmic scale factor.
The
first term in Eq.(\ref{eq:boltz}) describes the annihilation process of CHAMPs
(and anti-CHAMPs) and $-Zgf^{2}_{\gamma}$
describes all possible production
 processes. The parameter $Z$ characterizes the electromagnetic annihilation
 process and is given by
\begin{equation}
Z = 2M<\sigma v > m_{Pl} \sqrt{\frac{45}{16\pi ^{3} N_{eff}}}~~~.
\label{eq:z}
\end{equation}
$ < \sigma v > $ is an average annihilation cross section, which we assume to
have the form
\begin{equation}
 < \sigma v > \simeq \pi N^{4}\alpha ^{2}/M^{2}
\simeq \pi / \alpha ^{2}M^{2}~~~,
\end{equation}
 where $M \simeq Nm_{*}$ or $N \simeq 1/\alpha$.
If the particles whose reactions produce CHAMPs are in thermal
equilibrium themselves, it was shown by Turner \cite{turner} that
for $x<1$, $g(x)$ is
given by an equilibrium distribution, i.e.
\begin{equation}
g(x) \simeq \frac{1}{x^{3}}\exp (- \frac{2}{x}) ~~~.
\end{equation}
If we
neglect the annihilation term in Eq.(\ref{eq:boltz}) we obtain
\begin{equation}
\frac{d}{dx}\left(\frac{n_{M}(T)}{T^{3}}\right) =
 -a\frac{1}{x^{3}}\exp(-\frac{2}{x})~~~,
\end{equation}
where  $a = {\cal O}(10)$ if $M = {\cal O}(1/\sqrt{\alpha})m_{Pl}$ , i.e.
\begin{equation}
\frac{n_{M}(T_{*})}{T^{3}_{*}} = \frac{a}{2}\exp(-\frac{2}{x_{i}})
\left(\frac{1}{x_{i}} + 0.5 \right) +
\frac{n_{M}(T_{i})}{T^{3}_{i}}~~~,
\end{equation}
if $T_{*} << T_{i}$.
Here $x_{i}=T_{i}/M$. This solution actually overestimates $n_M(T)$, since
we have neglected the annihilation process. However, by inspection we now
see that the annihilation term can be neglected in comparison with the
production term in eq.(\ref{eq:boltz}).
If the  initial temperature $T_{i}$ is such that
$x_{i} \approx 0.028$ and the initial
density fulfills $n_{M}(T_{i}) = 0$, then we would get at $T = T_{*}$,
e.g. the today temperature of the cosmic background radiation,
a relic density compatible with the cosmological critical mass
density $\rho _{c}$. Since we should require $T_{i} \leq T_{Pl}$,
which is
easily fulfilled for the CHAMPs under consideration, we would then
obtain an unobservable small relic density for heavier CHAMPs, corresponding
to $M \gg m_{Pl}/\sqrt{\alpha}$, unless the
initial abundance $n_M(T_i)$ differs from zero and is such that
 $n_{M}(T_{i})/n_{\gamma}(T_{i})
= (n_{M}(T_{i})/n_{\gamma}(T_{i}))_{crit} \geq
{\cal O}(10^{-28})$.
This critical abundance corresponds to
$\rho _{c}$. We conclude that CHAMPs with
$M \simeq m_{Pl}/\sqrt{\alpha}$,
 could be thermally produced in the very early universe with no initial
 abundance
and thereby leading to a critical density.

If the initial abundance is much larger than
$(n_{M}(T_{i})/n_{\gamma}(T_{i}))_{crit}$ and $m_{*}^{2}/m^{2}_{pl} \simeq
 {\cal O}(\alpha )$, we expect
the freeze-out temperature $T_{d}$ to be below the
Planck scale within a few orders of magnitude. One can then imagine a
completely
different scenario for the cosmic production
of superheavy CHAMPs as compared to the analysis of Ref.\cite{glas}.
A mechanism of diluting
an {\it early} matter dominated phase of the
universe has been discussed in quite a detail by
Polnarev and Khlopov \cite{pol}. We will thus not discuss
that scenario in great detail. We would like, however, to point out that the
most simple aspects of this scenario may lead to an upper bound on $m_{*}$,
which is not in contradiction with the requirement that  $\alpha \leq
 (m_{*}/m_{Pl})^{2}$.
The physical picture is again an early radiation dominated phase with
subsequent thermal production of CHAMPs.
Let $\nu _{d} = \nu (T_{d})$, where $\nu (T) =n_{M}(T)/n_{r}(T)$, be the ratio
 of the
number densities of CHAMPs ($n_{M}$) and relativistic degrees of freedom
($n_{r}$)
at the freeze out temperature $T=T_{d}$. The universe will then develop into a
 matter dominated phase at the temperature \cite{pol}
\begin{equation}
T \simeq \nu _{d} M~~~,
\end{equation}
where we demand that $\nu _{d}$ is much smaller than one. Small initial metric
 perturbations  can grow large
in this
matter dominated phase and thereby
convert the primordial gas of CHAMPs and anti-CHAMPs into primordial
black holes (PBH), provided the early
matter dominated phase lasts sufficiently long.
(PBHs with a mass $\leq 10^{15} g$
would have been evaporated
today by Hawking radiation. The time-scale $\tau $ for such an
evaporation process
is $\tau \simeq 10^{10}years(M_{BH}/10^{15})^{3}$,
where $M_{BH}$ is the mass of the black hole in units of $g$.) In this scenario
the
early matter dominated stage is assumed to end by a gradual
transition into a radiation dominated phase.
Besides the formation of PBHs,
it may be that a certain amount of CHAMPs
survive.
Their abundance will
then, of course, be bounded by the critical density $\rho _{c}$. The
production of
PBHs will, however, put limits on $M$ and hence on $m_{*}$, due
to the observed
spectrum of PBHs
\cite{pol,carr}. For this scenario to work it requires that the relic abundance
of CHAMPs at $T= T_{d}$ does not dominate the energy density of the universe:
$n_{M}M \leq n_{r}T_{d} \leq n_{r}T_{Pl}$, i.e. $\nu _{d} M/m_{Pl} \leq
1$. The Boltzmann equation,
which determines the
temperature $T_{d}$ at decoupling and the relic abundance of charged boson
particles with mass $M$, is \cite{wolfram}
\begin{equation}
\frac{df(x)}{dx} = Z\left( f^{2}(x) - f_{eq}^{2}(x)\right)~~~,
\label{eq:boltzmann}
\end{equation}
with $x=T/M$ and $Z$ is again given by Eq.(\ref{eq:z}).
Here we have, although this is  questionable,
allowed for an initial condition
corresponding to thermal equilibrium in order to get an estimate of $f(x_{d})$,
where $x_{d} = T_{d}/M$.

In the expression for $Z$, $<\sigma v>$
is an
average annihilation cross-section, where for the relative particle
anti-particle
velocity $v$ we insert a typical virial velocity taken from the
PBH-formation process
and which is of the form $v \simeq \delta _{0}^{1/2}$~\cite{pol}.
Here the initial metric perturbation $\delta (\tilde{M})$ on a mass-scale
 $\tilde{M}$ is parametrized by  $\delta _{0}$ and $n$, with
\begin{equation}
\delta (\tilde{M})= \delta _{0}
\left( \frac{\tilde{M}}{\tilde{M}_{0}} \right)^{-n}~~~,
\label{eq:deltametric}
\end{equation}
where $\tilde{M}_{0}$ is the mass inside the horizon at the beginning of the
early matter dominated phase.
The annihilation cross
section for
the superheavy CHAMPs is estimated similarly as for the
magnetic monopoles, i.e. the dominant process is the emission of dipole
radiation. Thus we get \cite{zelkhlo}
\begin{equation}
<\sigma v> \simeq \pi \alpha ^{2} \frac{v^{-9/5}}{M^{2}}N^{4}~~~.
\end{equation}
The equilibrium distribution $f_{eq}(x)$ in Eq.(\ref{eq:boltzmann}) is
now
\begin{equation}
f_{eq}(x) = \frac{1}{2\pi ^{2}} \int_{0}^{\infty} dy
 \frac{y^{2}}{\exp(\sqrt{y^{2} + 1/x^{2}}) -1}~~~.
\end{equation}
The freeze-out temperature $T_{d}$ corresponds to the temperature when
the expansion rate $R^{-1}(t)dR(t)/dt$ is comparable to the reaction rate
$<\sigma v>n_{eq}$. This temperature can
approximatively be determined by considering
\begin{equation}
\frac{df_{eq}(x_{d})}{dx_{d}} = Zf^{2}_{eq}(x_{d})~~~.
\label{eq:td}
\end{equation}
For $x_{d}\ll1$, Eq.(\ref{eq:td}) leads to  $1/x_{d} + 0.5\log(1/x_{d}) \simeq
\
log (Z/(2\pi)^{3/2})$.
We then find that
\begin{equation}
f(x_{d}) \simeq \sqrt{2}f_{eq}(x_{d}) \simeq \frac{\sqrt{2}}{Zx_{d}^{2}}~~~.
\end{equation}
 For metric perturbations in the range
${\cal O}(10^{-5})\leq \delta _{0} \leq {\cal O}(10^{-3})$ we can use
 $1/x_{d} \simeq {\cal O}(15)$. The bound $\nu _{d} M/m_{Pl} \leq 1$
leads then, within an order of magnitude, to
\begin{equation}
\frac{1}{\alpha}\left(\frac{m_{*}}{m_{Pl}}\right) \leq \frac{1}{\alpha
 ^{2}}\delta _{0}^{-0.9}x_{d}^{2}~~~.
\label{eq:bound}
\end{equation}
In Fig.9 we present the numerical
evaluation of the Boltzmann
equation Eq.(\ref{eq:boltzmann}) in the range $0.001
 \leq x \leq 0.5$ where, for our purpose, we consider as initial data
 $f(x_{i})=f_{eq}(x_{i})$ with $x_{i}=0.5$ and $Z = {\cal O}(10^{6})$. The
 numerical value of $f(x_{d}) \simeq {\cal O}(10^{-5})$ does not depend
strongly on the actual value of $x_{i}$. The results obtained numerically are
in good agreement with the qualitative solution presented above.
Notice that the bound, Eq.(\ref{eq:bound}), is not in contradiction
with the condition $\alpha \leq (m_{*}/m_{Pl})^{2}$.
For a scale invariant initial metric perturbation, i.e. $n=0$ in
 Eq.(\ref{eq:deltametric}), with
$\delta _{0} \simeq 10^{-3}$ and following the analysis
for magnetic monopoles of Ref.\cite{pol} one would get $M \leq 10^{17} GeV$.
As pointed out in Ref.\cite{pol} such a limit
does, however,  not apply if the initial metric perturbation
is to small, i.e. if $\delta _{0} \leq  (\delta _{0})_{min} = 6\times 10^{-4}$.
The recent {\it COBE} data \cite{cobe} give $\delta _{0}
\simeq
{\cal O} (10^{-5})$ and thus we would not get such a strong bound on $m_*$ in
 this way. The scenario discussed above, therefore, suggests that an early
matter dominated phase can be diluted by the process of primordial
black hole formation. We are then lead to the thermal production scenario
with a small initial abundance, in which case we have argued that, for
the CHAMPs under consideration, the today cosmic abundance will be
small but nevertheless may lead to a critical density.

Since the CHAMPs we considered are superheavy, they will not affect
big-bang
nucleosynthesis. A likely scenario is that the compact charged boson stars
formed will
not bind to nuclei as is the case for the CHAMPs discussed in Ref.\cite{glas}.
Thus at least a fraction of the dark matter of the universe could be
made of compact boson stars.

As discussed in the previous section there is a
limiting charge for such objects, i.e. $Z
\leq 0.5\alpha ^{-1}$. With regard
to possible detection of such a dark matter candidate we notice that compact
bose stars with large $Z$ will,
with regard to their ionization properties, behave like superheavy magnetic
monopoles. Supermassive
electrically charged particles have recently been looked for by making use of
plastic track detectors sensitive to masses $M \geq 10^{-7} m_{Pl}$
\cite{orito}. The
bound on their number density, $n_{M}$, relative to the number density of the
cosmic background radiation
$n_{\gamma} \simeq  400(T_{\gamma}/2.7 K)^{3} cm^{-3}$ was found to be
$n_{M}/n_{\gamma} \leq 10^{-29}$.
Even such a small relative fraction of supermassive CHAMPs can lead to a
critical density for the universe, as we already pointed out above.
It is e.g. sufficient to consider compact charged boson stars with a mass $M
\simeq 250 m_{Pl}$.  Another completely different scenario of diluting an early
 abundance is, of course, inflation which we have not considered here.

Strongly interacting microgram dark matter has  been discussed
recently \cite{walsh}. For that case one may
replace in the above calculation the annihilation
cross-section with a suitably scaled strong-interaction
annihilation cross-section as in
Ref.\cite{glas}.
The previous analysis can be repeated for such
CHAMPs with a resulting less restrictive bound on $m_*$.
If the physical meaning of the $U(1)$ charge of the CHAMPs under consideration
is different from the electromagnetic coupling, then alternative
astrophysical scenarios may be possible. This has been discussed by Madsen and
Liddle \cite{madsen}.
\vspace{3mm}
\begin{center} {\bf ACKNOWLEDGEMENT}
\end{center}
\vspace{3mm}
We are grateful to John Ellis and
the CERN TH division for hospitality during a period when the present work
was initiated and A. D. Linde for many useful discussions.
P. J. wishes to thank the Institute
of Theoretical Physics at Chalmers University of
Technology for hospitality. B.-S. S.
wishes to thank the Institute of Theoretical
Physics at University of Z\"{u}rich for
hospitality.
%
\section*{\centering Figure Captions}

{\bf Fig.1} The metric functions $A(\tilde r)$ and $B(\tilde r)$ (upper and
 lower curve
 respectively in Fig.1a), the scalar field $\tilde {\phi} _{0}(\tilde r)$
 and the electric field $E(\tilde r)$ (Fig.1b)
as a function of $\tilde {r} = m_{*}r$ for a typical ``star"
with $\tilde {\phi}_0 (0)=0.1$, $\tilde e ^{2}=0.4$, $M\approx 166m_{*}$ and
$N \approx 168$.

\noindent {\bf Fig.2} The effective potential $V^{D+}_{eff}$, upper curve,
($V^{D-}_{eff}$, lower curve)
for electrons (positrons) with mass $m_{e}$, i.e. $m_{e}/m_{*}
 \simeq 0$,  as
 a function of $\tilde {r} = m_{*}r$
for a positively charged ``star" with $\tilde{\phi} _{0}(0) =0.1$, $\tilde e
 ^{2}=0.4$
 and $Q \approx 168e$ (Fig.2a).
For a sufficiently large value of  $\tilde {r}$, $V^{D+}_{eff}$ becomes
negative
. This signals that the bound states dive into
the negative continuum of the Dirac sea and that the vacuum becomes unstable
against $e^{+}e^{-}$-pairs production. In Fig.2b we show the corresponding
ef\-fec\-ti\-ve potentials if $m_{e} = m_{*}$.
In this case the bound states will not
dive into the negative continuum. ($V_{eff}^{D+}$: upper curve;
$V_{eff}^{D-}$: lower curve.)

\noindent {\bf Fig.3} The mass $\tilde M$
of the charged boson star in units of
$m^{2}_{Pl}/m_{*}$ as a function of $\tilde{\phi}_{0} (0)$ for various values
of the effective charge $\tilde{e}$. The line (the dashed one) going through
the maxima of the mass is drawn. Equilibrium configurations to the left of this
line are dynamically stable, whereas the other ones are unstable. The lines
above which the vacuum becomes unstable against pion pair production (dotted
line), which is determined by $Z=0.5/\alpha$,
and electron positron pair production (dotted dashed line) are also
drawn.

\noindent {\bf Fig.4} The lower branch of the curves is the
bound state spectrum of the Dirac equation for negatively
charged fermions
of mass $m_{*}$ in a $1s_{1/2}$ state ($\kappa = -1$)
in the field of a positively charged
boson star as a function of $\tilde{e}^2$ with
$\tilde\phi_0(0)=0.2$ (continuous line) and $\tilde\phi_0(0)=0.5$
(dashed line). The latter choice corresponds to
a dynamically unstable configuration. The upper branch of the curves
 correspond to the anti-fermion bound states. The
binding energy becomes zero nearby
the critical value of the $\tilde{e}$-coupling.
The corresponding spectrum
for the Klein-Gordon equation agrees, within the accuracy of our results,
with that of the Dirac equation.

\noindent {\bf Fig.5} The Klein-Gordon equation ground-state spectrum for
 negatively charged bosons
with mass $m$ in the field of a positively charged ``star"
approximated by a
uniformly charged sphere of radius $R$ and charge $Ze$. As $R$ tends
to zero the spectrum dives faster into the negative continuum corresponding to
 anti-particle states. In the point-like limit ($R \to 0$), we reach the
critical value $Z_{c}=0.5/\alpha$.

\noindent {\bf Fig.6}
 The effective potential $V^{K+}_{eff}$, upper curve,
($V^{K-}_{eff}$: lower curve)
for ``pions" (anti-``pions") with mass $m$, where we consider as an
 illustrative example $m/m_{*} = 0.05$,  as
 a function of $\tilde {r} = m_{*}r$
for a positively charged ``star" with $\tilde{\phi} _{0}(0) =0.1$, $\tilde e
 ^{2}=0.4$
 and $Q \approx 168e$ (Fig.6a).
$V^{K+}_{eff}$ is negative and less than -0.05 for sufficiently
small values of $\tilde {r}$.
This signals that the bound states dive into
the negative continuum of the
anti-``pions" and that the vacuum becomes unstable
against particle production. In Fig.6b we show the corresponding
effective potentials if $m = m_{*}$. In this case the bound states do not
dive into the negative continuum. ($V_{eff}^{K+}$: upper curve;
$V_{eff}^{K-}$: lower curve.)

\noindent {\bf Fig.7}
The particle number density $
n(\tilde r) = \frac{4\pi}{e}J^{0}(\tilde r)\tilde r^{2}$, such that
$N = \int _{0}^{\infty} n(\tilde r)d\tilde r$, for
a dynamically stable
boson star with $\tilde {\phi}_0 (0)=0.2$, $\tilde e ^{2}=0.25$
corresponding to
$N \approx 66$ in the presence of an induced charged condensate. The solid
curve is the number density of the star itself . The dashed curve
corresponds to the particle number density of the condensate (multiplied with a
factor $50$) in which we use $m^{2}/m_{*}^{2} = 10^{-3}$
as an illustrative example. The mean number of
particles in the condensate is  $N_{c} \approx 10$. With regard to the
gravitational properties of the boson star the back-reaction
of the condensate on
the star is negligible. On the scale of the figure, the particle number
density of the star in the absence of the condensate actually coincides
with the solid curve.

\noindent {\bf Fig. 8} The effective charge of the bose ``star",
 $Z_{eff}=
Z-q$, where $-q$ is the total charge of the pion condensate. The ``star"
is approximated by a positive
uniformly charged sphere with radius $R$ and charge $Ze$. The critical
 charge
$Z_{c}$ of the ``star" is determined for fixed $R$ by finding $Z$ such that
the binding energy $E=-m_e$, which is very close to zero
so that it can practically be taken equal to zero.
As $R$ tends to zero the screening becomes
more efficient.

\noindent {\bf Fig. 9} The number
density $n_{M}(T)$ of charged boson stars with
$M \simeq m_{Pl}/\sqrt{\alpha}$
(solid line) as derived from the Boltzmann equation
valid for a radiation dominated universe for the range $0.001<x<0.5$ and
$Z={\cal O}(10^{6})$. The initial data is such that
$n_{M}(T_{i}) = n_{eq}(T_{i})$, where we chose $T_{i}/M = 0.5$. As a comparison
we also plot (dashed line) $n_{eq}(T_{i})/T^{3}_{i} \approx 0.053$, which
corresponds to the solution of the Boltzmann equation
for charged bosons with mass $M=m_{*}
\simeq m_{Pl} \sqrt{\alpha}$
and an initial equilibrium distribution, i.e. in this case the
bosons remain in thermal equilibrium.
\end{document}